\documentclass[manuscript]{acmart}
\AtBeginDocument{%
  \providecommand\BibTeX{{%
    \normalfont B\kern-0.5em{\scshape i\kern-0.25em b}\kern-0.8em\TeX}}}

\setcopyright{acmcopyright}
\copyrightyear{2024}
\acmYear{2024}
\acmDOI{XXXXXXX.XXXXXXX}

\usepackage{pgfplots}
\usepackage{booktabs}

\begin{document}

\title{The Evolution of Emojis for Sharing Emotions: A Systematic Review of the HCI Literature}

\author{Charles Chiang}
\affiliation{%
  \institution{University of Notre Dame}
  \streetaddress{247 Fitzpatrick Hall}
  \city{South Bend}
  \state{Indiana}
  \country{United States}}
\email{cchiang3@nd.edu}

\author{Diego Gomez-Zara}
\affiliation{%
  \institution{University of Notre Dame}
  \streetaddress{353 Fitzpatrick Hall}
  \city{South Bend}
  \state{Indiana}
  \country{United States}}
\email{dgomezara@nd.edu}


\begin{abstract}
  With the prevalence of instant messaging and social media platforms, emojis have become important artifacts for expressing emotions and feelings in our daily lives. We ask how HCI researchers have examined the role and evolution of emojis in sharing emotions over the past 10 years. We conducted a systematic literature review of papers addressing emojis employed for emotion communication between users. After screening more than 1,000 articles, we identified 42 articles of studies analyzing ways and systems that enable users to share emotions with emojis. Two main themes described how these papers have (1) improved how users select the right emoji from an increasing emoji lexicon, and (2) employed emojis in new ways and digital materials to enhance communication. We also discovered an increasingly broad scope of functionality across appearance, medium, and affordance. We discuss and offer insights into potential opportunities and challenges emojis will bring for HCI research.
\end{abstract}



\received{12 September 2024}

\maketitle

\section{Introduction}
With the prevalence of instant messaging and social media platforms, emojis have become important artifacts for expressing emotions and feelings in computer-mediated communication. According to the Unicode Consortium, 92\% of the world's online population uses emojis \cite{greg_emoji_nodate}. In Adobe Research's' 2022 U.S. Emoji Trend Report, 91\% of respondents said emojis make it easier to express themselves, and 75\% of them felt more connected to people who use emojis \cite{fonts_team_future_2022}. While emojis began as a set of simple icons in the 1990s, they experienced a rapid expansion in variety and complexity due to the advancement of smartphones and the growing number of social media platforms in the world \cite{noauthor_faq_2023}. Furthermore, the increasing personalization and incorporation of multimedia components have transformed emojis into new interactive objects, such as stickers, avatars, and memes \cite{konrad_sticker_2020}. As such, emojis have significantly transformed and enhanced online communication over the past decades, and become a cultural staple in and of itself \cite{alonso_emoji_2017}.

One of the main reasons for emoji's popularity is their capability to provide nonverbal markers of emotion to text messages \cite{wiseman_repurposing_2018, kelly_characterising_2015}. Previous research has shown that emojis can strengthen the meaning of a message, helping readers better understand its sentiment by providing visual and emotional context \cite{chairunnisa_analysis_2017}. Their designs and pictorial elements enable nonverbal cues from text and enable users to express emotions outside of words \cite{gawne_emoji_2019, novak_sentiment_2015, boutet_emojis_2021}. As a result, users can convey different tones and emotions that might be lost or hard to express using plain text \cite{highfield_instagrammatics_2016}. Furthermore, emojis are used in highly personalized ways, where they may take on multiple functions depending on the relationship between their users \cite{wiseman_repurposing_2018}. People can make their own meaning out of emojis to create a ``shared uniqueness,'' where an emoji can completely change its meaning depending on who the emoji is sent to, as well as the sender's relationship with that person. Emojis have also become more customizable, allowing users to select different skin tones, genders, and body characteristics to present themselves more authentically \cite{robertson_self-representation_2018}. Because of these customizable and non-textual properties, social media platforms and messaging apps have adapted emojis for creating stickers and reactions, increasing their already widespread ubiquity as well as providing an effort-free way to use them \cite{seta_biaoqing_2018,konrad_sticker_2020}. Emojis are now capable of eliciting joy among users and supporting more meaningful communications across a broader range of spaces \cite{chen_exploring_2017}.

Despite the popular and ubiquitous use of emojis in modern online communication, there is still a gap in the literature surrounding the way emojis have evolved. As emojis continue to shape online communication between millions of users, HCI researchers are called to understand how emojis' unique roles, designs, and functions have evolved in online communication systems over the past years \cite{bai_systematic_2019}. Mapping the existing research on emojis can help HCI researchers transform emoji's design and application, as well as envision more effective, engaging, and interactive online communication systems. Prior literature has identified different ways in which people, platforms, and systems employ emojis, taking a broader view of emoji to include usage as communication markers, pictures, and even as \textit{lingua franca} for worldwide users \cite{alshenqeeti_are_2016, gawne_emoji_2019, maier_emojis_2023, dresner_functions_2010}. As platforms push forward with novel introductions to the emoji space, aiming to garner the attention of users as emoji originally did, they begin to differ wildly from traditional emojis \cite{bai_systematic_2019, seta_biaoqing_2018}. However, these systems are still primarily used to fill a similar role to emojis---to express emotions, clarify intents, and for fun \cite{tang_emoticon_2019}. Therefore, examining the emojis' visual appearance, functionality, and mode of communication will be fundamental to understanding their design, development, and use in the future. In this paper, we seek to answer the following research questions:
\begin{itemize}
    \item \textbf{RQ1: How have emojis evolved to better communicate users' emotions?}
    \item \textbf{RQ2: How do the changes in emojis affect their ability to communicate emotion?}
\end{itemize} 

To answer the research questions, we conducted a systematic literature review to map how the HCI community has studied, designed, and employed emojis for sharing emotions in online communication. After screening over 1,100 articles from the ACM Digital Library, IEEExplore, and Web of Science, we identified 42 relevant articles that described methods, systems, models, and studies employing emojis to share emotions, reactions, and feelings. From this corpus, we conducted a thematic analysis to identify enduring themes that described the findings and contributions of these articles. We found that our corpus revolved around two different themes: (1) improving how users find and select emojis to represent their emotions and (2) employing emojis in new ways and digital materials to enhance their communication. The first theme included methods to improve recommendations, input selection, and emojis themselves. The second theme included enhancements and new characteristics of emojis to expand their communicative space.

Our contributions are as follows. First, we provide a comprehensive review of the HCI literature on emojis for sharing emotions in computer-mediated communication. We focus on their function as tools for facilitating communication between different people and their evolution over the past few years. We present our results, which indicate that emojis' designs and purposes have drastically changed over time, as researchers seek to expand the domain of emojis through different mediums of communication, ways of usage, and means of emotion conveyance. We highlight an increasing design divergence within the themes generated from our corpus, with one branch continuing to further improve how users interact with modern emoji, and the other branch exploring more experimental methods to connect user meaning with emoji communicative function. Finally, we discuss and offer insights into the potential opportunities and challenges that emojis will bring for future HCI research. 

\section{Background}
In this section, we review relevant studies covering emotions in computer-mediated communication (CMC). We then revise the evolution of emojis and their use, including describing their increasing affordances. Lastly, we conclude this section with our research question.

\subsection{Sharing Emotions in Computer-Mediated Communication}
One of the main challenges in online communication is overcoming the lack of non-verbal cues \cite{archer_words_1977}. While in face-to-face (F2F) communication individuals can employ nonverbal cues that significantly contribute to the understanding of their message, such as facial, tonal, and physical expressions, those elements are missing in online communication. This is due to the tools and adaptations that communicators have developed in response to the inherent lack of nonverbal cues in CMC. One such adaptation is making nonverbal cues verbal by expressing their intent directly \cite{walther_let_2005}.

According to media richness theories, the effectiveness of online communication technologies relies on the amount and quality of transmitted information \cite{ishii_revisiting_2019}. While F2F interactions allow individuals to use their bodies, other materials, and the space to communicate, the same information cannot be transmitted through low-richness mediums such as documents, notes, or memos. People prefer either lean or rich communication channels depending on the media sender's relation and motives \cite{kwak_self-disclosure_2012}, as there are benefits and drawbacks to both. While the most common reason that people choose F2F over CMC is because of the ability to use nonverbal cues, the most common reason for using CMC over F2F is to transmit information remotely and shield themself from the message recipient \cite{riordan_emotion_2010}.

In the context of CMC, the most well-researched tools for conveying emotion are \textit{emoticons} and \textit{emojis} \cite{riordan_emojis_2017}. They can be used as nonverbal markers of emotion, skepticism, politeness, and more within CMC communication \cite{vandergriff_emotive_2013}. The use of such icons for visualizing emotions enables `leaner' ways to communicate emotions using CMC, approaching the `richer' expressiveness of F2F communication \cite{ahn_emoticons_2011, derks_role_2008}. 

\subsection{Emojis Evolution}
Emojis are defined as colorful and cartoon pictographs embedded in text messages posted on websites and online devices \cite{noauthor_utr_2016}. The word `emoji' comes from the Japanese words picture (`e') and written character (`moji') \cite{noauthor_utr_2016}. They were initially created to provide visual representations and fill in emotional cues that are hard to express or convey using only text \cite{miller_blissfully_2016}. Over the past three decades, computer science researchers developed solutions to enable representations of these nonverbal cues in textual communication. One of the earlier forms was `emoticons,' which are letters and punctuation marks arranged within the text to represent an expressive face (e.g., :D ) \cite{bai_systematic_2019}. Nass posited that emoticons provide a mechanism to transmit emotion when individuals do not have voice \cite{noauthor_digital_2007}. By the end of the 1990s, the first set of dedicated pictographs was released by Japanese phone carrier \textit{SoftBank} \cite{burge_correcting_2019}. While this is thought to be the first set of emoji, it was not until Shigetaka Kurita created a set of 176 emojis for DoCoMo's `i--mode' in 1999 that emoji began to see widespread use. The popularity of i--mode emojis led other platforms worldwide (including MSN Messenger, BlackBerry, and Apple) to create their own emoji sets, each slightly different than the other \cite{evers_msn_2003, alonso_emoji_2017}. 

While unique emoji sets across platforms served to draw engagement to their respective messaging platforms, the lack of cross--platform standardization meant that emoji was restricted from broader internet usage. Recognizing the need for standardization, there was a proposal as early as 2000 to encode the i--mode emojis into Unicode \cite{noauthor_utr_2016}. In 2006, Google worked on converting the various Japanese emojis into Unicode, which continued through 2007. Finally, in 2010, the Unicode Consortium accepted the proposal to adopt 625 emoji characters. Since then, the Unicode Consortium has been responsible for decisions on new emoji characters as well as generally ensuring their equivalence across platforms \cite{noauthor_unicode_2024}. Modern emojis are characterized by a defined set of agreed-upon expressions, which are determined by the Unicode Consortium. According to \textit{Emojipedia}, emojis appear differently across platforms since the ``artwork is specific to which fonts are included on the system'' \cite{noauthor_faq_nodate}. This may reflect different design philosophies and artistic styles, as well as satisfy customers' needs for uniqueness to draw them to their platform as opposed to competitors \cite{abosag_customers_2020}.

The integration of emojis into major tech platforms increased their worldwide popularity. In 2011, Apple added an emoji keyboard to its mobile operative system (iOS), making it easy for iPhone users to access and use emojis \cite{alonso_emoji_2017}. Google followed this decision and integrated emojis into Android \cite{alonso_emoji_2017}. The widespread use of these operating systems helped propel emojis into mainstream usage. Moreover, social media platforms like Facebook, Twitter, WhatsApp, and Instagram adopted and promoted emojis \cite{alonso_emoji_2017}.

In more recent years, scholars have discussed the evolution of emoji into other forms of communication artifacts, most notably \textit{stickers} \cite{konrad_sticker_2020}. Stickers are pictograph ``images, usually larger than graphical emoticons and emojis, offered as thematic sets in the communication interfaces of instant messaging apps and social networking services, often organized in tabs and personalized collections'' \cite{seta_biaoqing_2018}. While emojis can be used in conjunction with text or other media providing context \cite{mcculloch_linguistic_nodate}, stickers can only be sent as the sole message, with no additional context provided within the message \cite{bai_systematic_2019}. Moreover, stickers are highly contextual and are not fully designed to convey emotions. While emojis are simple and generalizable tools to share emotions, stickers can transmit other kinds of information, such as jokes, memes, impressions, or animations that require users' social context to be understood \cite{wang_culturally-embedded_2019, cha_complex_2018}. In addition, stickers are tied directly to the platform that the sender is on. While similar stickers may exist across platforms, they lack the intrinsic ubiquity of emoji \cite{lee_smiley_2016}. The unfettered access to creating and using stickers, alongside the fact that stickers are unique to each platform, has led to a greater diversity of stickers overall. As of 2024, there are less than 4,000 total emojis, while back in 2017, there were nearly 500 thousand individual stickers on one messaging app (iMessage) alone \cite{list_stickerstats_2017}.

\subsection{Transmitting emotions through emojis}
To understand the role of emojis in communicating emotions, we build on Derks et al. \cite{derks_role_2008}'s \textit{emotion communication} framework. According to this framework, transmitting emotions through CMC faces reduces social presence when expressing emotions and reduces visibility when recognizing emotions. As a consequence, people can feel more comfortable expressing more intense and frequent positive and negative emotions when compared to F2F.

Researchers have extensively discussed the effectiveness of emojis to communicate emotions. Alshenqeeti \cite{alshenqeeti_are_2016} shows that emojis convey tone, intent, and feelings to fill their role as nonverbal cues. The nonverbal information that emoji conveys in CMC has been likened to physical gestures \cite{gawne_emoji_2019} and facial expressions \cite{grosz_semantics_2023} in F2F communication. These nonverbal cues reduce ambiguity as well as alter the emotional intensity within digital messages \cite{archer_words_1977, lee_effect_2002}. Meanwhile, Dresner et al. \cite{dresner_functions_2010} highlight the ability of emoji to communicate pragmatically, without necessarily attaching emotion. As a `pragmatic indicator', emojis can be used to replace words or ideas in a sentence without changing the emotional context \cite{dresner_functions_2010}. The effect of emoji in a message depends heavily on the context in which it is used, how it is used, and how it is interpreted by the reader \cite{alshenqeeti_are_2016, dresner_functions_2010}.

In sum, HCI researchers have delved into solutions and experimented with diverse systems to support and examine the effectiveness of communicating emotions through emojis. Consequently, our goal is to investigate how emojis have been employed to facilitate this purpose over the past years. By conducting a systematic literature review, we aim to delineate the solutions, insights, and recommendations that emojis have already provided and could offer to advance CMC communication further.

\section{Methodology}
To answer our research questions, we followed a scoping review methodology \cite{arksey_scoping_2005,levac_scoping_2010} to synthesize existing research and identify trends in research employing emojis. This methodology allows researchers to map the literature and identify gaps in a specific area of research. Scoping reviews are transparent, comprehensive, less prone to bias, and make it easier to reproduce the detailed information reported about each step of the review and how it is conducted \cite{harris_joining_2019}. They have gained prominence in the HCI community as they allow synthesizing and comprehending the development of different research directions \cite{shibuya_mapping_2022, rogers_much_2022, hirzle_when_2023, cosio_virtual_2023}. We used the Preferred Reporting Items for Systematic Reviews and Meta-Analyses (PRISMA) guidelines to report the methods and results \cite{page_prisma_2021}. PRISMA covers four stages in the review process: Identification, Screening, Eligibility, and Inclusion. We collected papers and tabulated them in a shared Google Spreadsheet, capturing metadata such as publication year, abstract, and keywords. Once the metadata was recorded, we hid the articles' authors to avoid potential bias during the coding phase.

\subsection{Eligibility Criteria for Inclusion of Articles}
\emph{Inclusion.} This review contains articles that pertain to emojis or systems centered on emojis. Articles were eligible for inclusion if they focused on emojis as they pertain to communicating, expressing, or sharing emotions. We included papers that were written starting in 2004 as we want to focus on developments in emoji use.


\emph{Exclusion.} We excluded articles that were written in a language other than English or could not be downloaded from their websites. We also excluded articles that did not study any form of communicating emotions, moods, or feelings. We removed studies that only analyzed current norms of emoji usage and perception, rather than their communicative purposes. We decided not to include derivate terms from emojis, such as ``sticker,'' ``avatars,'' or ``emoticons,'' to keep our focus exclusively on the evolution of emojis. Lastly, we removed meta-analysis or literature review articles as our goal was to analyze research articles. 

\subsection{Data Sources}
We chose three data sources highly employed by HCI researchers: the ACM Digital Library, the IEEE Xplore Digital Library, and the Web of Science database (WoS). We built search queries to find papers examining emojis as a form of non-verbal/non-textual communication for emotion. We iterated through several rounds of search queries before deciding which to use. The final search query was: \textit{``emotion'' AND ``communication'' AND ``emoji'' }. We chose these search terms as they allowed us to find emoji-related studies that focus on how to communicate emotions through emojis. The same search query was used in all three databases. 

We conducted the search in June 2024 and gathered 1,122 articles through the ACM Digital Library, 80 through the IEEE Xplore Digital Library, and 187 through the Web of Science. In total, we collected 1,389 articles that included duplicates. From this search, only two articles were published before 2014. The other 1,120 articles from ACM DL were published from 2015 onwards. We exported the results of the IEEE Xplore and Web of Science searches to CSV files, and the results of the ACM Digital Library search were exported to BibTeX as citations through their website. This BibTeX file was then converted to a CSV file. We then tabulated the articles into a Google Spreadsheet. We then removed 120 duplicates using the articles' DOIs, resulting in 1,149 articles.

\subsection{Article Selection}
The first author (henceforth referred to as the coder) manually screened all retrieved articles according to the Inclusion and Exclusion criteria in a three-stage process. In the first stage, the coder reviewed the articles' titles and determined their eligibility for inclusion according to the criteria previously set. This produced 203 articles that were deemed eligible. In the second stage, the coder reviewed the articles' abstracts in addition to their titles. If the coder found that the articles met the eligibility criteria, then they were retained. After the second stage of review, 72 articles were selected. In the third stage, the coder reviewed the full text of the eligible articles in order to determine eligibility. The coder then reviewed all selections made and finalized the corpus. In total, 42 articles were deemed eligible.

\subsection{Data Extraction and Synthesis}
Once the final corpus was established, the coder extracted data to synthesize their characteristics, methods, and findings. Using a Google Spreadsheet, the coder extracted data pertaining to:

\begin{itemize}
    \item Year of publication
    \item Type of System's Innovation (e.g., using biosignals, using voice)
    \item Article Type: Research paper, short paper (four pages or fewer), and posters.
    \item Evaluation type: User Study, Deployment Study, Benchmark Testing
    \item Methodology: Quantitative, Qualitative, or Mixed Methods. 
\end{itemize}

Using these 42 articles, we conducted a thematic analysis to create cohesive themes that described this final corpus \cite{braun_using_2006}. The goal of thematic analysis is to discover themes from the data while minimizing confirmation bias. We began by familiarizing ourselves with the articles and taking notes on the main contributions of the paper, as well as the ideas mentioned throughout. These notes were compiled in a Google spreadsheet and used to inductively generate codes without trying to fit these codes into preexisting coding frameworks or theories \cite{nowell_thematic_2017,braun_using_2006}. The coder generated codes by revising each paper and used these codes to recognize patterns throughout the corpus. The codes were not mutually exclusive since the articles could address multiple codes. The coder iterated the initial codes through several rounds of adjustments until no new ideas emerged from them. Following this, the codes were collected into themes, which also underwent iterative rounds of renaming to create more accurate descriptions. Through several meetings, all authors discussed the themes to ensure they both agreed upon the definition and boundaries of each theme. After several iterations, the authors arrived at the three crucial themes that will be discussed in the following sections.

\section{Results}
\begin{figure}
  \includegraphics[width=0.8\textwidth]{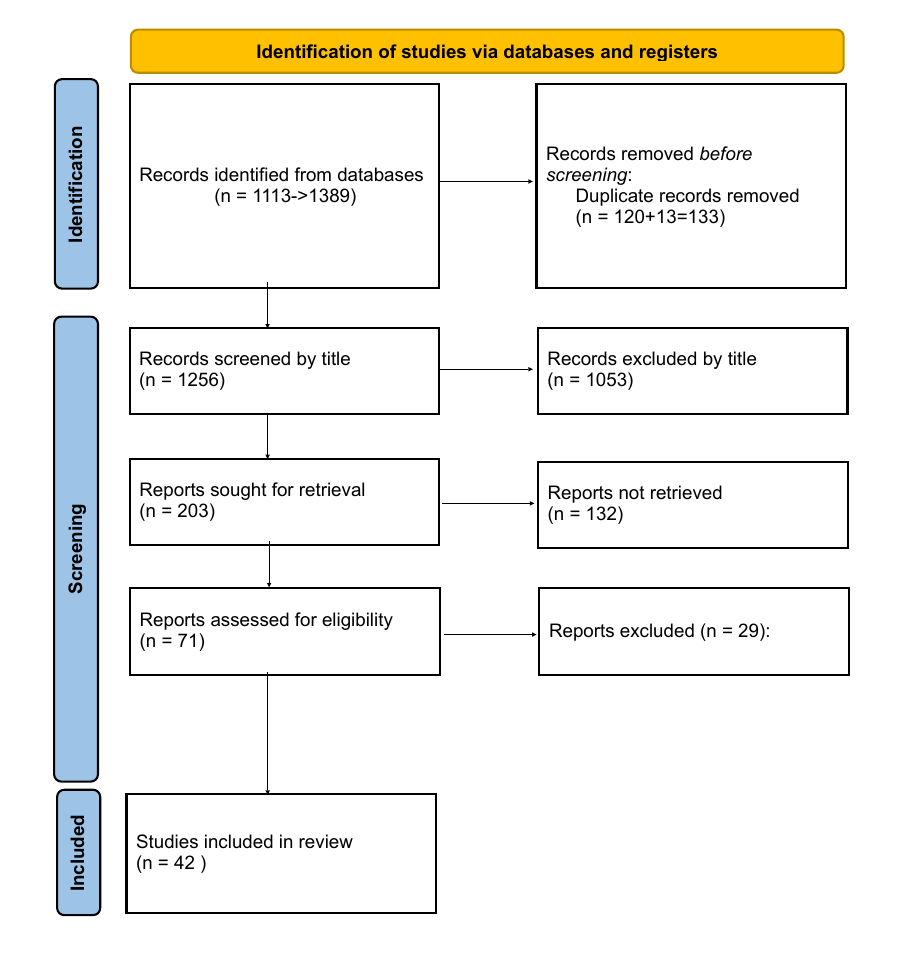}
  \caption{PRISMA Flow Diagram showing the article selection process}
  \Description{PRISMA Flow Diagram showing the article selection process}
  \label{fig:prisma}
\end{figure}

Our final corpus considers 42 articles that addressed methods and systems employing emojis to communicate and share emotions among users. Figure \ref{fig:prisma} shows the filtering process of the papers in IEEE Xplore, ACM, and Web of Science databases. After screening more than 1,300 articles, the most frequent reason to exclude articles in the screening stage was that they did not focus on enabling users to communicate their emotions using emojis. Of the records assessed for eligibility, 20 passed through the initial screening but were not included in the final review. Many of the excluded articles did not present a system or design pertaining to communicating emotions using emojis.

\subsection{Description of the Articles}
\begin{figure}
\begin{tikzpicture}
\begin{axis}[x tick label style={
		/pgf/number format/1000 sep=},
  ybar interval, ymax=8,ymin=0, minor y tick num = 1]
\addplot coordinates { (2016, 1) (2017, 6) (2018, 4) (2019, 7) (2020, 6) (2021, 5) (2022, 4) (2023, 6) (2024, 3)};
\end{axis}
\end{tikzpicture}
\caption{Publication years of the articles}
\label{fig1}
\end{figure}
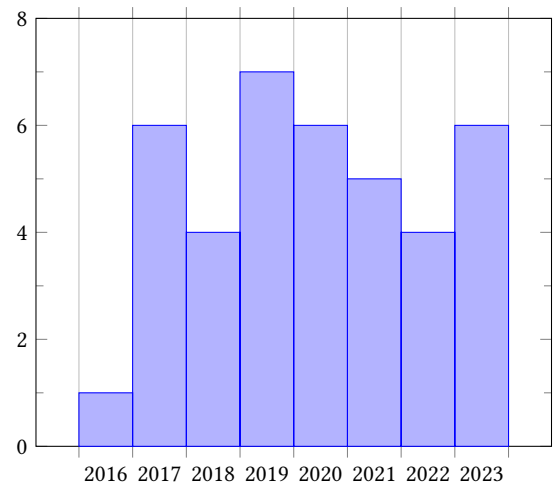

Figure \ref{fig1} shows the publication dates of the articles included in the review. Most of the articles were published in the ACM Digital Library (n=36), followed by four articles published in the Web of Science database, and two articles in the IEEE Digital Library. All of the reviewed papers were published from 2016 onwards. We hypothesize that there were no papers from before 2016 included in the final corpus due to the lack of extensive emoji research, which has been steadily increasing over the past 10 years. Out of the 42 articles reviewed, 30 were evaluated through user studies or deployment studies. Of those that did not include a user evaluation, six were short papers, three were posters, one contributed a theoretical framework, and one contributed a gesture set. Five articles included benchmark comparisons or some method of verifying model performance. Of the 30 papers that performed user studies, ten involved a deployment study (ranging from one to four weeks) while one article was deployed as a full application. Twenty-nine papers had users participate by themselves (13) or in pairs (16). Three of these articles were lenient on the group size, and allowed a group of three. Two articles studied groups of three or more. Thirteen articles encouraged participants to sign up as pairs or groups, and one study randomly assigned participants to pairs. These results are also shown in Table \ref{table:1}.

\begin{table}[!htb]
\small
\begin{tabular}{llllll}
\hline
\multicolumn{2}{l}{\textbf{Characteristic}}         & \textbf{Number} & \textbf{Percentage} &  &  \\ \hline
\multicolumn{2}{l}{\textbf{Source}}                 &                 &                     &  &  \\
          & ACM Digital Library                     & 36              & 86\%                &  &  \\
          & Web of Science                          & 4               & 10\%                &  &  \\
          & IEEE Xplore Digital Library             & 2               & 5\%                 &  &  \\
\multicolumn{2}{l}{\textbf{Years of Publication}}   &                 &                     &  &  \\
          & 2016-2018                               & 11              & 26\%                &  &  \\
          & 2019-2021                               & 18              & 43\%                &  &  \\
          & 2022-2024                               & 13              & 31\%                &  &  \\
\multicolumn{2}{l}{\textbf{Type of Article}}        &                 &                     &  &  \\
          & Research Paper                          & 30              & 71\%                &  &  \\
          & Short Paper                             & 9               & 21\%                &  &  \\
          & Poster                                  & 3               & 7\%                 &  &  \\ \hline
\multicolumn{2}{l}{\textbf{Type of Evaluation}}     &                 &                     &  &  \\
          & User Study                              & 19              & 45\%                &  &  \\
          & Deployment Study                        & 10              & 24\%                &  &  \\
          & User Study and Deployment Study         & 1               & 2\%                 &  &  \\
          & Other                                   & 5               & 12\%                &  &  \\
          & None                                    & 7               & 17\%                &  &  \\
\multicolumn{2}{l}{\textbf{Evaluation Methodology}} &                 &                     &  &  \\
          & Mixed Methods                           & 17              & 40\%                &  &  \\
          & Qualitative                             & 9               & 21\%                &  &  \\
          & Quantitative                            & 9               & 21\%                &  &  \\
          & N/A                                     & 7               & 17\%                &  &  \\
\multicolumn{2}{l}{\textbf{\# of Users}}            &                 &                     &  &  \\
          & 1                                       & 13              & 31\%                &  &  \\
          & 2                                       & 13              & 31\%                &  &  \\
          & 2-3                                     & 3               & 7\%                 &  &  \\
          & 3+                                      & 2               & 5\%                 &  &  \\
          & N/A                                     & 11              & 26\%                &  & 
\end{tabular}
\caption{Characteristics of the Articles}
\label{table:1}
\end{table}

\subsection{Article Classification}
The following article classifications emerged from the thematic analysis described above. We divided the papers into two main thematic categories: ``Emotion Discovery with Emojis Made Simple'' and ``Augmenting emojis' affordances and characteristics.'' We also further identify sub-categories based on how these themes are implemented within the papers. Papers could belong to one or more categories as they addressed several of these themes. Details are listed in Table \ref{table:2}.

\subsubsection{Emotion Discovery with Emojis Made Simple}
This theme describes how systems enable users to determine and chose emojis that best express their emotions. The articles include research on input methods to find emojis, implementing changes to existing emojis, and creating new emojis in real time. We found 17 articles on this theme, which encompasses the following sub-themes: connecting users to emojis, guiding emoji recommendation, and emoji expressiveness. 

\paragraph{Connecting Users to Emojis ($n=10$).} With the continuous addition of new emojis, the list has expanded significantly from its original size and makes it harder for users to find the appropiate emoji to communicate their emotions. To alleviate some of these design limitations, the articles of this sub-theme introduce alternative and improved methods to identify emojis based on the user's desired emotion. For example, Pohl et al. \cite{pohl_emojizoom_2016} presented one of the first articles to research emoji entry: a novel keyboard built around zooming rather than scrolling. This design allows for exploratory interaction in two dimensions rather than one, utilizing spatial memory more effectively and allowing users to enter emojis up to 18\% faster. In a subsequent article, Pohl et al. \cite{pohl_beyond_2017} presented two further emoji keyboard models based on semantic and annotated information, while simultaneously highlighting the need for further development in emoji entry. Rathod et al. presented a system for recommending emojis by predicting emoji usage based on the context of the conversation, including to whom the speaker is messaging \cite{rathod_emoji_2023}. This explores the findings from Wiseman et al. \cite{wiseman_repurposing_2018} of personalized communication through emojis. Research also shows more multimodal alternatives to keyboard-based emoji entry, such as video and voice. Zhang et al. \cite{zhang_voicemoji_2021} presented an application allowing visually impaired users to issue verbal commands that correspond to relevant emojis. Koh et al. \cite{koh_developing_2019} created a hand gesture set and recognition system that maps to emojis. Other systems use facial recognition to map user's displayed emotions to emojis \cite{liu_reactionbot_2018, henriques_emotionally-aware_2018, liu_integration_2019}. These systems aim to reduce the mental effort required in the technical usage of emojis, which can help users select emojis that better represent their emotions \cite{pohl_beyond_2017, henriques_emotionally-aware_2018}.

\paragraph{Enriching Expressions through Emojis ($n=3$).} These articles presented research on new formats, images, or icons that expand the emotional scope of emoji. Originally, emojis were conceived as simple and pretty recognizable symbols that resemble objects or emotions from everyday life. However, emojis were not originally designed with the explicit intent to be as expressive as possible. Articles from this corpus reveal that the set of emojis has gradually become more nuanced with the addition of new emojis and allowing users to customize pre-existing emojis. More expressive emojis could benefit users with visual impairments or difficulty in recognizing emotions. Many articles of this theme focused on increasing the accessibility of emojis, either in recognizing the emotions they convey or for those users with disabilities. Cherbonnier et al. \cite{cherbonnier_recognition_2021} designed a new set of emojis that conveyed six basic emotions with greater intensity and recognition when compared to Facebook emojis. In one of the studies they performed, participants were able to recognize emotions conveyed by the 'new emoticons' more accurately and successfully than other emojis or even images of faces. Further studies confirmed this advantage but emphasized it for certain emojis, primarily disgust and sadness. Choi et al. \cite{choi_image-based_2020}, introduced tactile emojis for visually impaired individuals that were recognized correctly 81\% of the time and were able to improve the clarity of the sentence. This study shows that image-based tactile emojis could be beneficial for visually impaired individuals, and begin to minimize the gap in the communication environment between sighted and visually impaired individuals. Finally, Nishimori et al. presented a method to customize the expressiveness of an emoji on the fly \cite{nishimori_--fly_2023}. Users can swipe up or down on an emoji to change some accessory elements on the emoji (such as the size of a heart, or tears), which in turn alters the emotional expressiveness of the emoji.

\paragraph{Guiding Emoji Recommendation ($n=4$).} One simple way to recommend emojis is by replacing words or emoticons in a text conversation with their corresponding emojis (e.g. the word \textit{fire} becoming a fire emoji). While this is suitable for emoji and emoticon's function as a literal pictorial representation of words, it falls flat when it comes to more complex interpretations of emojis. Articles in this category focused on contributions towards enhancing these recommendations to users. These articles used mostly natural language processing and machine learning techniques to improve recommendations' accuracy and variety of emojis. Guibon et al. \cite{guibon_emoji_2018} introduced a model using sentiment-related features, allowing them to predict the emoji used with an 84\% F1 score and 95\% precision in a corpus of private instant messages. Kim and Gong et al. \cite{kim_no_2020} presented a model that uses similar features across a larger scope of the chat, allowing for more context across multiple sentences. This allowed users to choose emojis up to 38\% faster than baseline and reportedly more suited to the conversation. In another study, Hong et al. \cite{hong_moji_2024} presented a keyboard interface using query expansion and emoji prediction to better suggest emojis for users. Lastly, Gao et al. \cite{gao_learning_2020} introduced a similar model for sticker recommendation, learning representations without labels and considering context across multiple turns of dialog. The model achieved state-of-the-art performance across all metrics and paved the way for a personalized sticker response selection system.

\subsubsection{Augmenting emojis' affordances and characteristics}
Within the corpus, we see some papers whose contributions deviate greatly from the traditional definition of emoji. However, the authors of these papers use emojis as a reference or a comparison to their systems. We include these papers because they are important to understanding the current state of HCI literature in emotion communication. The second emerging theme was focused on studies and systems that combined emerging technologies with emojis in innovative ways, drastically diverging from their common usage. It included 20 articles that introduced new properties, affordances, and characteristics to emojis in order to enhance their ability to communicate emotions. The sub-categories are: (1) Connecting the Body and Emojis, (2) Employing Augmented Reality (AR) with Emoji, (3) Color and Shape, (4) Emoji in Context, and (5) Leveraging Physical Space. One article \cite{buschek_personal_2018} was classified into three sub-categories, as the authors studied three different systems. Two other articles \cite{chen_bubble_2021, semertzidis_neo-noumena_2020} were classified into two sub-categories, as their implications are relevant for both.

\paragraph{Emojis as Artifacts of Users' Data ($n=3$)} 
As some of the systems identified in this corpus have allowed users to customize the search and use of emojis more than before, emojis are inherently embedding more user information, such as their identity, relationships, and emotions. Many of the papers reviewed in our corpus utilize data from the user in order to facilitate their goals, be it text data from the conversation, audio/visual data, or raw biosignal data. This trend can potentially contain information about users' appearance, location, and potentially even physical state. This information is not only transferred through the emoji but also recorded by the emoji within the context of the conversation and the user. Several articles in the review presented some form of automatic emotion detection, such as image capturing or facial recognition. Across these papers, many participants voiced some negative sentiment toward the automatic function. For example, Liu et al. \cite{liu_reactionbot_2018} presented a system that automatically attaches reactions to Slack messages by detecting facial expressions. While this enhanced the genuine expression of emotion, it also led some to worry about their portrayal of emotion. Participants showed concern about leaking emotion. Furthermore, contrary to their hypothesis, ReactionBot reduced the social presence felt by users. Lee et al. \cite{lee_exploring_2023} created `ARWand', a system to enable users to create asynchronous augmented reality (AR) messages. `ARWand' featured an automatic reaction-capturing video, which helped users be more authentic but also led to privacy concerns about their surroundings. Poguntke et al. \cite{poguntke_smile_2019} created four visualizations, representing emotions captured automatically by a webcam. The automaticity of the reactions highlighted users' desire to maintain control over their emotional reactions online. This connects with the hypothesis posed by Derks et al. \cite{derks_role_2008}, which states that the reduced spontaneity typically associated with CMC allows for more regulation of emotions.

\paragraph{Connecting the User's Body with Emojis $(n=11)$.} This theme focuses on furthering the connection between the users' bodies and the emotions they want to portray through emojis and other visualizations. The articles within this category use a variety of means to achieve this effect. Some systems transform physical bio-information into emotions through the use of existing biosensory technologies (such as smartwatches) or by supplementing text messaging with data from extraneous sensors. Liu et al. \cite{liu_supporting_2017} and Buschek et al. \cite{buschek_personal_2018} investigated the effects of sharing heart rate with a partner alongside text messages, finding that the information was expressive and promoted mindfulness of their own as well as their partners' heart rate. Expanding on this work, Liu et al. \cite{liu_animo_2019} created a smartwatch app with a shareable set of animations that played according to the emotional state of the wearer, serving as a lightweight social connection tool. These animations were highly abstract, consisting of a shape, color, and motion that changes according to the biosignal data. In another study, Liu et al. \cite{liu_significant_2021} used an animated otter, and found that allowing the animations to 'sense' biosignals enabled a more genuine emotional connection than just the animation without biosignal sensing. Other papers discussed how brain activity visualization can also affect interpersonal emotions. In Liu et al. \cite{liu_can_2017}, different visualizations of brain waves and the associated emotional state were compared. Participants were willing to use the expressive biosignals to form impressions about others. Semertzidis et al. \cite{semertzidis_neo-noumena_2020} introduces a brain-computer interface that utilizes artificial intelligence to read users' emotional state. Their system, Neo-Noumena, used emotional valence and arousal to create 3D fractals and dynamically display them to their partners in mixed reality. Using the system resulted in significant improvement in measures of emotion regulation and challenged users' perceptions of emotion representation.

Other papers classified in this theme explored how vibrations, heat, and other haptic motions can communicate emotions like emojis do. Haptic feedback is widely used to draw users' attention to incoming information or notifications. While vibrations are the most common, other modalities of haptic feedback are less commonplace. The articles within this theme expanded upon the different types of haptic feedback, and combined it with text-based messaging to replace emoji's role in nonverbal emotion communication. Wilson et al. \cite{wilson_multi-moji_2017} systematically combined multiple different haptic modalities across three studies (vibrotactile, thermal, and vibrotactile + thermal) to show that the combination of vibrotactile, thermal, and visual increases the affective emotional range of messages the greatest. An et al. \cite{an_vibemoji_2022} developed VibEmoji, offering users a system to create multi-modal emojis using animation and vibration. Participants reported that the animation and vibrotactile effects enhanced the meanings of emojis, and even helped them create new meanings associated with the emojis. Participants also found that the emojis were able to set and reset the atmosphere of the conversation. Araujo de Aguiar et al. \cite{araujo_de_aguiar_touch_2023} proposes a system that incorporates light, vibrations, tactile interactions, and a digital display with the aim of improving emotion communication. The system combines a wearable device to transfer multi-modal haptic feedback with an app that enables text messaging as well as some social media.

Finally, other papers have used audio to augment the expression of emotions through emojis. Voice messaging provides an alternative to text messaging, allowing users to utilize voice inflections and sound to convey their message with a greater affect. These articles explored different ways to add emotions to voice messages. Chen et al. \cite{chen_bubble_2021} studied how coloring the background of a voice message had an intensifying effect on emotional arousal. The coloring of the background was based on the emotion conveyed in the voice message, and was found to intensify emotions on average. However, users were less willing to use the different colors in the case of negative emotions. In another example, Haas et al. \cite{haas_voicemessage_2020} created a system that allows users to augment voice messages with other audio such as background sounds and voice changers. Users found the system more expressive and personal compared to traditional voice messages. Extending previous work (VibEmoji \cite{an_vibemoji_2022}, discussed above), An et al. \cite{an_emowear_2024} created a system that allows users to choose emotional teasers on voice messages, that significantly improved the communication experience for both senders and receivers.

\paragraph{Employing Augmented Reality (AR) with Emojis $(n=7)$.} These articles explored ways in which AR technologies can employ emojis to overlay reality and communicate emotion more intimately. Namikawa et al. \cite{namikawa_emojicam_2021} proposed the design of an AR system that covers users' faces with emojis over video calls, either automatically or through manual input. This system allowed users to control emotional delivery more accurately. This design also allowed users to read the emotions of the speaker and audience members more easily. In another study, Zhang et al. \cite{zhang_auggie_2022} proposed a digital avatar that can be dressed, posed, animated, and sent as a personal handcrafted message. Their system, ``Auggie,'' highlights personal efforts while minimizing procedural efforts in order to allow users to create meaningful and personal messages in a lightweight format. Lee et al. \cite{lee_exploring_2023} explored this idea more, prototyping a messaging system centered around AR messages and capturing video reactions. Participants found that AR messages allowed for communication and connection in an immersive and ``fun'' way. Leong et al. \cite{leong_wemoji_2022} presented an app for using AR effects in a shared physical space to add another dimension to the conversation. This process also enabled the creation of design reflections for future mixed reality emoji systems. Bhatia et al. borrowed visual language from comics to investigate the different effects on emotion and sensations \cite{bhatia_using_2024}. The comic annotations, which are visual effects derived from comics, were found to enhance the emotion associated with an action or object similar to how emojis influence the meanings of a sentence. Though this area is still in its infancy, the common theme reported is that AR may be a novel way to interact more immersively with each other. Semertzidis et al. \cite{semertzidis_neo-noumena_2020} introduced their system, Neo-Noumena, which used VR as a way to actualize the emotions portrayed by users in space and time. Users found the system was able to augment their communication, not just facilitate it. Finally, Sun et al. \cite{sun_anon-emoji_2019} presented a system utilizing optical see-through, overlaying participants' facial expressions with emoji. The goal of the system was to help understand how children with autism spectrum disorders perceive emotions.

\paragraph{Color and Shape $(n=5)$.} These five articles studied how modifying the colors and shapes of text bubbles has an effect on communicating emotions, in comparison and conjunction with emojis. In particular, alterations were made to the outlined form and background color of text bubbles as well as other objects within the interface in order to produce an emotional effect. For example, Buschek et al. \cite{buschek_personal_2018} studied the effects of various changes to the style, size, and form of text, and concluded that this allowed for more recognizable and individual typing compared to standard. The system studied, TapScript, included a emoji drawing feature that participants used to emphasize the personal expression offered by the system. Aoki et al. \cite{aoki_emoballoon_2022} created a system that used speech input to create text with a corresponding speech balloon with different shape according to the intensity of emotion. This study suggested that using these emotional speech balloons results in a lesser misunderstanding between senders' and receivers' emotional arousal when compared to using emojis. An et al. \cite{an_affective_2023} presents ``AniBalloons,'' which imbues chat balloons with affective animations, designed with frequently used emojis in mind. Participants responses emphasized the ability of the ``AniBalloons'' to be an additional source of emotion expression alongside emojis, as well as being more abstract and less personalized when compared to emojis. As discussed above, Chen et al. \cite{chen_bubble_2021} studied how coloring the background of a voice message had an intensifying effect on emotional arousal. The background color intensified the emotional effect in the `excited,' `sad,' and `angry' emotions. However, the `serene' emotion was unaffected. Lastly, Bhatia et al. \cite{bhatia_using_2024} utilized elements from comics, such as speed or scent lines to emphasize different sensations. Users reported having more positive emotions associated with the elements presented in the article.

\paragraph{Emoji and Context $(n=2)$.} In this sub-theme, two articles discussed and examined how leveraging additional users' contextual information can enhance the use of emojis to share emotions. Their goal is to transmit more information about the senders' context that is unknown to the recipients. Buschek et al. \cite{buschek_personal_2018} studied the effects of contextual information such as weather, location, and activity being added to a text conversation, comparing this information to emoji's role in providing additional context to a message. This direct contextual information reflects the authors' design principle of minimizing ambiguity in interpretation. Through three systems and studies, the paper found that users were able to infer better the meaning of the messages by incorporating additional contextual information. These studies also emphasized the importance of facilitating shared understanding between users. In particular, Khandekar et al. \cite{khandekar_opico_2019} created an emoji-only social media site that utilized location-based posts. Through a full deployment, surveys, and interviews, their system ``Opico'' showed that though messages comprised solely of emoji required more mental effort to interpret than straightforward text, there were benefits in the speed of sending messages, users' enjoyment, and the greater amount of creativity expressed. Users strongly relied on the contextual data of location and time to decipher the meaning of the emoji messages.

\begin{table}[!htb]
\begin{tabular}{@{}ll@{}}
\toprule
                           & \textbf{Included Articles} \\ \midrule
\textbf{Improvement}       &                            \\
\textit{Connecting Users to Emoji}             & \cite{alvina_mojiboard_2019, el_ali_face2emoji_2017, henriques_emotionally-aware_2018, koh_developing_2019, liu_integration_2019, liu_reactionbot_2018, olshevsky_touchless_2021, pohl_beyond_2017, pohl_emojizoom_2016, zhang_voicemoji_2021}                           \\

\textit{Enriching Emoji through Expression}        & \cite{choi_image-based_2020, cherbonnier_recognition_2021, nishimori_--fly_2023}                    \\ 
\textit{Guiding Emoji Recommendation}     & \cite{kim_no_2020, gao_learning_2020, guibon_emoji_2018, hong_moji_2024},                         \\ \midrule
\textbf{Enhancement}       &                            \\
\textit{Emojis as Artifacts of Users' Data} & \cite{poguntke_smile_2019, liu_reactionbot_2018, lee_exploring_2023} \\
\textit{Connecting the Body with Emojis}        & \cite{semertzidis_neo-noumena_2020, liu_significant_2021, jiang_intimasea_2023, buschek_personal_2018, liu_animo_2019, liu_can_2017, liu_supporting_2017, haas_voicemessage_2020, chen_bubble_2021, an_vibemoji_2022, wilson_multi-moji_2017, an_emowear_2024, araujo_de_aguiar_touch_2023}
                           \\

\textit{Employing Augmenting Reality with Emoji} &  \cite{leong_wemoji_2022, zhang_auggie_2022, lee_exploring_2023, sun_anon-emoji_2019, semertzidis_neo-noumena_2020, namikawa_emojicam_2021, bhatia_using_2024}                         \\
\textit{Color and Shape}   &  \cite{chen_bubble_2021, aoki_emoballoon_2022, buschek_personal_2018, bhatia_using_2024, an_affective_2023}                         \\
\textit{Emoji in Context}        &  \cite{khandekar_opico_2019, buschek_personal_2018}                                                   \\ \bottomrule
\end{tabular}
    \caption{Classification results}
    \label{table:2}
\end{table}

\section{Discussion}
In this study, we explored how the HCI literature has addressed the evolution of emojis for communicating emotions. After identifying 42 papers that examined, employed, and extended the use of emojis, we identified relevant themes that described how emojis' affordances, design, and capabilities have evolved over the past decade. These articles have introduced new ways for users to communicate their emotions through emojis, which have been constantly reshaped over the past years. In the following subsections, we elaborate on the key findings of this systematic literature review, focusing on how these articles delineate the use and design of emojis for sharing emotions, identifying gaps in this corpus, and outlining potential future research directions and opportunities.

\subsection{RQ1: How have emojis evolved to better communicate users' emotions?}
The corpus shows that emojis have evolved from text-messaging icons to more complex multimodal forms. These modalities include AR/VR/MR, biosignals, haptics, and speech bubble form. These subthemes have little to do with emoji as Unicode-based icons, and more to do with the role in which emoji plays as a nonverbal communicative marker of emotion. Notably, the identified articles frequently made direct comparisons to or cited emojis as inspiration for the emotion communication systems presented. In this way, we posit that the articles discussed represent the next step in the evolution of emojis for emotion communication.

In contrast to previous emoji studies, the corpus suggests that the scope of emojis has broadened, and emphasizes the increased affordances of emoji research. In more established systems, we can see a divergence in vocabulary following the different affordances provided, from emoticons and emojis to stickers, avatars, and reactions. Within the selected articles, however, the vocabulary is not yet divergent. These articles still represent their contributions in comparison to emoji --- systems such as Multi-Moji \cite{wilson_multi-moji_2017} and VibEmoji \cite{an_vibemoji_2022} clearly reference them in their title. Participants even compare some of the presented systems to using emojis \cite{liu_animo_2019}. Other articles directly present their systems as evolutions of emojis \cite{liu_significant_2021}. As the modalities discussed within the corpus gain popularity, the vocabulary associated with each theme may vary as well. This divergence may be indicative of a larger trend within online communication platforms and technologies, as users employ different systems, devices, and technologies to satisfy their communication needs \cite{gonzales_towards_2015}.

As Wiseman et al. \cite{wiseman_repurposing_2018} described, emojis can take on meaning outside of their direct visual interpretations. The selected articles shed light on further extensions of this meaning-making enabled by these different modalities. For example, `Auggie' \cite{zhang_auggie_2022}, which focused on creating extremely personalized, effortful experiences. Other articles employed biosignals to interpret users' emotions and generate specific emojis \cite{semertzidis_neo-noumena_2020, liu_supporting_2017, an_vibemoji_2022, liu_significant_2021}. The abstract visualizations did not deter participants and their communication partners from creating their own meaning \cite{zhang_auggie_2022, liu_significant_2021, liu_supporting_2017}. In fact, Liu et al. identifies this as an area where future work might focus on deliberately designing systems around subjective interpretations \cite{liu_can_2017}.

\begin{figure}
  \includegraphics[width=1.0\textwidth]{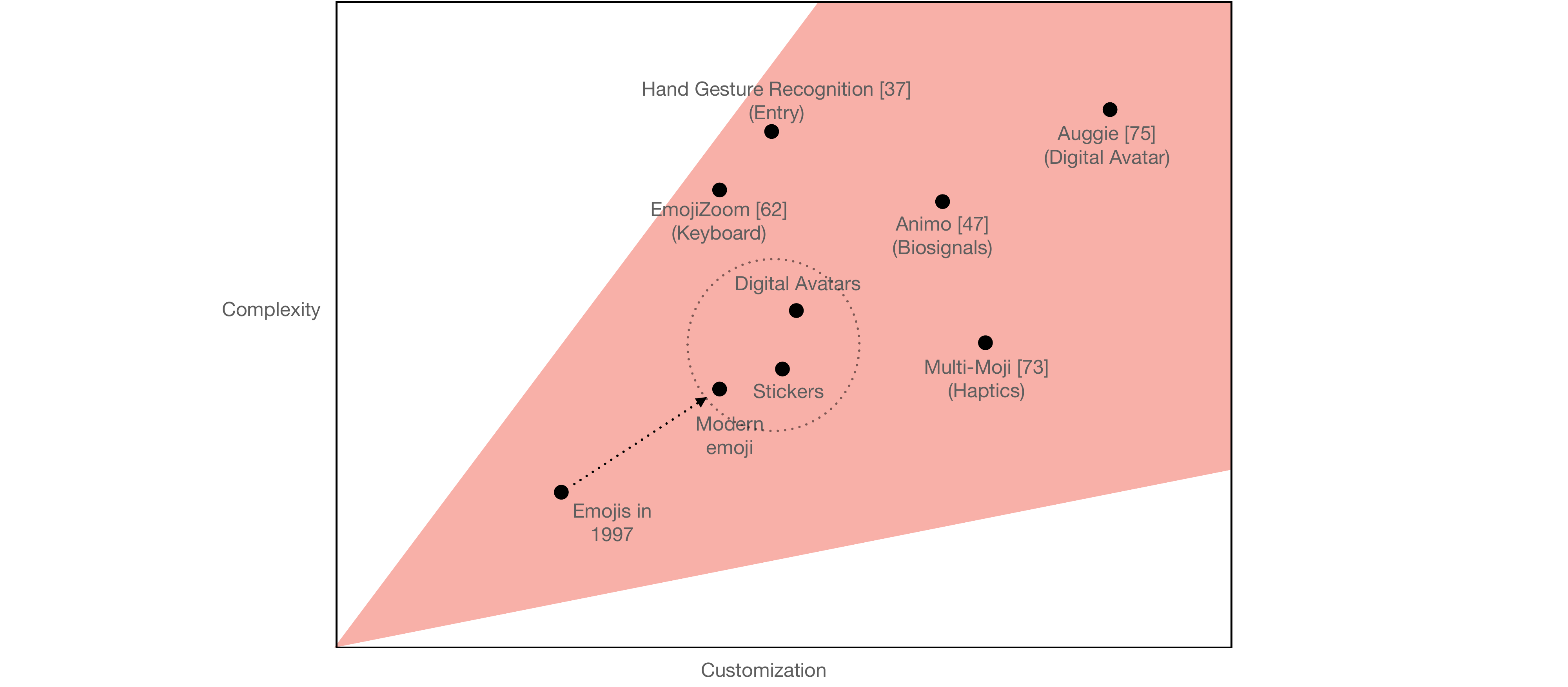}
  \caption{Customization and Complexity are positively correlated, but not necessarily linear}
  \Description{Customization and Complexity are positively correlated, but not necessarily linear}
  \label{fig:divergence}
\end{figure}

Finally, the articles suggest that there is a relationship between the amount of customization and the complexity of an emoji. The relationship between these two is not necessarily directly correlated. As a natural development of the technology, there is a trend towards greater customization for emotion communication. Mainstream emoji is still relatively limited in function compared to the systems presented in this literature review. The customization options of Unicode emoji are limited to visual appearance, and cross-platform emojis are not one-to-one copies of each other. The innovations presented in the corpus represent a variety of different affordances for emojis, and the customization --- complexity relationship reflects that as a trend broadly towards the upper right. While there is an associated amount of complexity as customization increases, the amount of effort put in does not necessarily need to increase linearly, as we depict in Figure \ref{fig:divergence}. Some articles deal with increasing or decreasing the effort required to use the system. A system that supports an easier way for users to send messages may reduce the mental burden associated with using it, and allow users to focus more on the content of the message being sent, such as by detecting cues from the body \cite{pohl_beyond_2017, liu_significant_2021, liu_supporting_2017}. However, a system that intentionally requires more effort may be a benefit, especially by emphasizing personal efforts for greater communicative affect \cite{zhang_auggie_2022}.

In summary, emojis have evolved into a diverse set of modalities to enhance their ability to communicate emotion. Research in AR/VR/MR, biosignals, and haptics represent the potential future of emojis. Though current vocabulary delineates emoji from these modalities, the work represented in corpus of articles suggests that this distinction may not exist in the future. Furthermore, we see a push towards designing for greater personal meaning and customization within emojis.

\subsection{RQ2: How do the changes in emojis affect their ability to communicate emotion?}
The corpus shows that emoji research has supported users' ability to create more personal, interpretable, and effective communications. The articles explore different levels of expressiveness over a broad range of emotions. The different modalities in which this research was conducted also had an effect on users' emotion communication. Moreover, the corpus suggests a trend in emoji research towards emphasizing the ability for users to create custom emojis. These changes show how emojis have become more personal and open to interpretation, similar to F2F communication.

The corpus suggests that emoji research has made emojis more interpretable for users through customization and contextual information. Emojis are traditionally icons depicting faces with an expression, which is relatively straightforward. Even so, there is still some amount of ambiguity surrounding emoji meanings, stemming from popular culture usage, context, and personalized meanings. The articles increase the contextual information provided alongside communication channels. Articles utilized data such as heart rate or input from voice and video to increase the information with which receivers can interpret the emotion communicated. Another way that articles were used is through the personal nature of communications. Abstract visualizations resulted in more ways for users to interpret the emotions, as long as they were consistent and not random. Similarly, customization of emojis or other communication objects allowed users to reframe their conversations depending on who the conversation was with. This emphasizes the ability for users to assign personal meanings to emojis depending on the conversation partner \cite{wiseman_repurposing_2018}.

The multiple modalities provided by emoji research have afforded users a greater communicative affect. With the variety of modalities discussed in the corpus, emoji research has afforded users more ways to convey information. Across the different modalities, users were additionally able to accurately convey emotion despite their relative unfamiliarity with the systems. Some studies investigated the differences between positive, negative, calm, or angry emotions, with varying willingness to use. Participants were more willing to portray positive emotions in some studies \cite{an_affective_2023, chen_bubble_2021}, and negative emotions in others \cite{semertzidis_neo-noumena_2020}.

Lastly, the corpus also provided evidence of how emoji research has allowed users to align their emotion communication intention with the emoji selection. Firstly, users can select an emoji that is more accurate to their emotions, such as by using facial recognition to map user's displayed emotions to emojis as seen in Liu et al. \cite{liu_reactionbot_2018} and Henriques et al. \cite{henriques_emotionally-aware_2018}. Secondly, users can select an appropriate emoji faster, such as through keyboard layouts such as EmojiZoom \cite{pohl_emojizoom_2016}. In fact, these ways are intertwined, and many articles discuss ways to improve both the accuracy of the emoji's emotion as well as speed, such as in recommendation systems \cite{kim_no_2020, gao_learning_2020, guibon_emoji_2018, hong_moji_2024} or reducing the mental effort required to select an emoji \cite{koh_developing_2019}. Finally, the set of emojis can be modified to have a greater communicative effect. Articles such as \cite{nishimori_--fly_2023} allow users to directly control the expressiveness displayed by an emoji. Additionally, as new emojis are added by the Unicode Consortium over time, there are more options for users to select appropriate emojis.

In summary, we found that emoji research has led users towards emotion communication that is similar to F2F communication. The new channels of communication are marked by ambiguity, interpretability, and personalization, characteristics shared by F2F communication. This challenges some established theories of media richness and CMC \cite{ishii_revisiting_2019}.

\subsection{Implications}

\subsubsection{Types of Relational Conversations}
Of the articles presented in the corpus, only some articles deal with users in groups rather than solo users. Due to emoji's heavily personalizable meaning, emphasizing dyadic relationships when evaluating work may be important. The articles discussed in the theme ``Augmenting emojis' affordances and characteristics'' are primarily concerned with more intimate conversations. This is likely due to the fact that they are more experimental and depart from established emojis and uses. The more bespoke systems are more difficult to deploy, and haptic / biosignal devices are not ubiquitous. The first theme, ``Emotion Discovery with Emojis Made Simple'' includes articles that deal with or could be extended to workplace environments, such as ReactionBot \cite{liu_reactionbot_2018}. This is likely due to the fact that it works around established norms of what emojis are, what they can do, and what they mean. However, this research is not specific to emotion communication or emoji in the context of the workplace or professional relationships. We identify that there is space for research to be done on the use of emojis for relationships between friends or colleagues.

\subsubsection{Automaticity and Recommendations}
The first category, "Emotion Discovery with Emoji" is focused on aligning users' emotions with the appropriate emoji. As discussed above, this alignment is a focus for emoji research. However, as emojis are regularly added by the Unicode Consortium, there becomes a greater diversity of expression available for users. The gradual increase in the set of emojis provides a sort of benchmark increase in the emotion alignment for users. Emoji research should aim to provide contributions other than adding new emojis to the set. It is also worth noting that a greater number of available emoji may come contrary to the goal of finding the appropriate emoji. This can be due to decision paralysis or the increasing time it takes to find the emoji on the keyboard. However, this again can be remedied through recommendation, which theoretically allows the system to align itself with the user's intended emotion and predict which emoji should be used. While recommendation algorithms are not relevant for every type of research work, designers should be aware of the findings and capabilities presented by such papers.

Across several articles in the corpus, many participants voiced some negative sentiment toward the automatic function. For example, Liu et al. \cite{liu_reactionbot_2018} presented a system that automatically attaches reactions to Slack messages by detecting facial expressions. While this enhanced the genuine expression of emotion, it also led some to worry about their portrayal of emotion. Participants showed concern about leaking emotion. Similarly, Lee et al. \cite{lee_exploring_2023} featured an automatic reaction-capturing video that led to privacy concerns, as participants may not want to share information about their surroundings. Even just the webcam requiring permissions in Poguntke et al. \cite{poguntke_smile_2019} was enough to give participants pause and an associated feeling of unease. This connects with the hypothesis posed by Derks et al. \cite{derks_role_2008}, which states that the reduced spontaneity typically associated with CMC allows for more regulation of emotions. Designers should be aware of the dangers of automaticity in CMC, and consider the benefits that CMC provides.

\subsubsection{Accessibility}
The affordances of emoji lend themselves more generally to sighted individuals, as they visually represent faces and icons. Within the corpus, we see articles that focus on making emojis more accessible to the visually impaired \cite{choi_image-based_2020}. Several articles \cite{zhang_voicemoji_2021} dealt with techniques that deliberately seek to enhance accessibility, either through voice entry \cite{zhang_voicemoji_2021} or for emotion recognition \cite{choi_image-based_2020}. However, this is a space that the corpus still is lacking in. Several articles introduce new modalities for emotion communication that are inaccessible to those outside of visual impairments. Effects that change colors in order to affect emotion may not work as well for those who are colorblind, and different entry techniques (i.e. swiping) or haptic feedback may be inaccessible for those with physical impairments. As more improvements are made towards the accessibility of emoji, designers should consider implications for those with different physical abilities outside of the visual.

While aimed at those with a disability, accessible research has the potential to benefit those without disability \cite{microsoft_wide_2003} In general, technologies designed for users have often proved to be beneficial for a wide range of users \cite{mott_smart_2016, wang_eartouch_2019}. Conversely, articles that do not have a focus on accessibility may prove to be effective for some, such as non-verbal emoji entry systems \cite{koh_developing_2019, el_ali_face2emoji_2017, liu_reactionbot_2018} or more affective emojis \cite{kim_messaging_2020}. Biosignals have been adopted for use in assisting those with disabilities in communicating with caretakers \cite{schultz_biosignal-based_2017, pinheiro_alternative_2011}, but not for more casual use. We identify this as a space for future work, to investigate how multiple modalities may be incorporated in designs for those with disabilities to communicate their emotions better.

\subsubsection{The Increasing Divergence Across CMC platforms}
Lastly, the increasing number of platforms and services is another main explanation for emojis' divergence. With the increasing number of social platforms competing for users' attention and experiences, designers have created more tailored emojis that respond to the specific experiences that these platforms aim to provide. Allowing users to use emojis to quickly react to other users' stories, as well as the choices of these emotions, proves that the future of emojis relies profoundly on how these platforms promote, curate, and trigger certain emotions. Future research should explore the consequences of this increasing partitioning and divergence of emojis across operative systems, platforms, and devices, as the choices will shape users' communication in the long term.

Furthermore, HCI designers and practitioners should discuss new ways to standardize these multiple families of emojis, as well as when they became part of the Unicode system. With standardization and consensus among emojis' purposes and characteristics, there will need to be a discussion about implementation, affordances, and visual congruity across platforms. Alternatively, if emojis do not converge into a few solutions, HCI designers may need to provide new ways to facilitate interoperability for these digital objects. For example, emojis created in one system such as ``Opico'' \cite{khandekar_opico_2019} cannot be used on Facebook. In contrast, Snapchat's `Bitmojis' can currently be sent as images in emails and SMS messages, which allows them to essentially be used on any platform that allows images to be sent. While future efforts to unify emojis among online platforms and systems will remain uncertain, characterizing the evolution of emojis can provide guidance for future considerations. 

The divergence in emoji research reflects a broader trend in the field of CMC technologies, where new tools and social platforms continually emerge and perish, each with unique features and designs. Understanding this divergence is essential for HCI researchers as they navigate the dynamic landscape of digital communication and strive to create more cohesive and user-friendly experiences. This is still a relatively unexplored space in HCI, and it is one that will continue to develop alongside the maturation of the technologies. While it is still too early to tell how emotion communication will be affected by the technologies of tomorrow, for now, we can be aware of some potential opportunities and challenges that will present themselves over time.

\subsection{Limitations and Future Work}
This study is not exempt from limitations. Firstly, excluding `stickers,' `avatars', or `emoticons' in our search query limited the potential scope of these articles. We decided not to include these terms as our research goal was to examine the evolution of emojis. Studies with stickers and emoticons that acknowledged emojis were still captured by our search query. Therefore, the terms used by our search query were sufficient for our research goal. Future systematic reviews should consider including these terms and also exploring how those pictograms have been employed to communicate emotions online. Secondly, one coder primarily conducted the eligibility and inclusion stages, which could have affected the reliability of the coding process. To mitigate the bias in this coding process, the authors met frequently to discuss the findings and selected papers. New themes and structures emerged from their conversations. We provide the spreadsheet with its notes and decisions as \textit{Supplemental Materials} to promote the reproducibility and analysis of our study. Lastly, systematic literature reviews are limited by the data sources and queries established by the research team. We did not include articles indexed in other research platforms (e.g., Google Scholar, Scopus), pre-printed versions (e.g., arXiv, SSRN), or non-academic outlets (e.g., blogs, industry reports). Including these datasets could provide more insights into the role of emojis in communicating emotions. 

\section{Conclusion}
We conducted a systematic literature review to study how HCI researchers have examined emojis to communicate emotions in the past 10 years. After identifying 42 articles, we found several themes motivated by both purpose and method, with the purpose of improving the search and selection of emojis, as well as enhancing emojis to provide new communicative affordances in a distinct manner. The literature provides different methods and systems that include improving recommendations, input selection, and emojis themselves. The review also shows how systems employ different signals and environments to enhance emojis' purposes and emotion communication. In conclusion, we hope this systematic review sheds light on the role of emojis for communicating emotions, providing HCI researchers and practitioners with discussions and insights into emojis' future affordances and components.

\bibliographystyle{ACM-Reference-Format}
\bibliography{main}

\end{document}